\documentclass[preprint,aps,showpacs,nofootinbib,preprintnumbers,amsmath,amssymb]{revtex4-1}
\usepackage{}
\usepackage{epsfig}
\usepackage{subfigure}
\usepackage{dcolumn}
\usepackage{bm}
\usepackage[usenames ,dvipsnames]{xcolor}
\usepackage{slashed}
\usepackage{graphicx,color}

\begin{document}
\title{ $f_J(2220)$ and Hadronic  $\bar B^0_s$ Decays}
\author{Y.K. Hsiao$^{1,2}$ and C.Q. Geng$^{1,2,3}$}
\affiliation{$^{1}$Physics Division, National Center for Theoretical Sciences, Hsinchu, Taiwan 300\\
$^{2}$Department of Physics, National Tsing Hua University, Hsinchu, Taiwan 300\\
$^{3}$College of Mathematics \& Physics, Chongqing University
of Posts \& Telecommunications, Chongqing, 400065, China}
\date{\today}

\begin{abstract}
We study the hadronic $\bar B^0_s$ decays based on the existence of the resonant state $f_J(2220)$.
In particular, we are able to explain
the unexpected large experimental result of ${\cal B}(\bar B^0_s\to J/\psi p\bar p)
=(3.0^{+1.2}_{-1.1}\pm 0.52\pm 0.03)\times 10^{-6}$ measured recently by the LHCb collaboration due to the resonant contribution
in $\bar B^0_s\to J/\psi f_J(2220)$ with $f_J(2220) \to p\bar p$,
while it is estimated to be at most of order $10^{-9}$ in terms of the OZI rule without the resonance.
In addition, we find that ${\cal B}(\bar B^0_s\to D^{*0}(f_J\to) p\bar p)=(4.70\pm 2.89)\times 10^{-7}$, 
${\cal B}(\bar B^0_s\to J/\psi(f_J\to)\pi\pi)=(15.6\pm 15.2)\times 10^{-6}$ and 
${\cal B}(\bar B^0_s\to D^{*0}(f_J\to)\pi\pi)=(24.5\pm 24.4)\times 10^{-7}$,
while
${\cal B}(\bar B^0_s\to J/\psi(f_J\to)K\bar K)<1.6\times 10^{-5}$ and
${\cal B}(\bar B^0_s\to D^{*0}(f_J\to)K\bar K)<2.5\times 10^{-6}$.
Moreover, we  predict that the decay branching ratios of 
$\bar B^0_s\to (J/\psi\,,D^{*0})\Lambda\bar \Lambda$ are $(2.68\pm1.23)\times 10^{-7}$ 
and $(2.25\pm0.80)\times 10^{-6}$. Some of the predicted $\bar B^0_s$ decays
are accessible to the experiments at the LHCb.
\end{abstract}

\maketitle
\section{introduction}
In some three-body $B$ meson decay of $B\to {\bf B\bar B'}M$,
with ${\bf B\bar B'}$ a baryon pair and $M$ a recoiled meson or photon,
the partial decay width as the function of  
$m_{\bf B\bar B'}=p_{\bf B}+p_{\bf \bar B'}$
is observed to have a peak near $m_{\bf B\bar B'}\simeq m_{\bf B}+m_{\bf \bar B'}$ of the threshold area.
This is the so-called threshold enhancement, which dominates the decay branching ratio 
of $B\to {\bf B\bar B'}M$. The examples of these decays include $B\to p\bar p M$ with $M=(D^{(*)},\,K^{(*)},\,\pi,\rho)$
and $B\to \Lambda\bar p M'$ with $M'=(\pi,\,\rho,\,\gamma)$. 
Theoretically, the threshold effect has been realized as the result of the perturbative QCD (pQCD) effect~\cite{countingrules1,countingrules2}.
Consequently,  many experimental data on the baryonic B decays can be well explained~\cite{KLLbar,angulardistribution,BtoJpsippbar}. 


However, it is not the case for $\bar B^0_s\to J/\psi p\bar p$.
 The branching ratio of $\bar B^0_s\to J/\psi p\bar p$ 
presented by the LHCb collaboration is given by~\cite{LHCb1}:
\begin{eqnarray}\label{data1}
{\cal B}(\bar B^0_s\to J/\psi p\bar p)
&=&(3.0^{+1.2}_{-1.1}\pm 0.52\pm 0.03)\times 10^{-6}\,,
\end{eqnarray}
where the first and second  uncertainties are
statistical and systematic, respectively, while the third one
originates from the control channel branching fraction measurement.
Note that ${\cal B}(\bar B^0\to J/\psi p\bar p)=(2.0^{+1.9}_{-1.7}\pm 0.9\pm 0.1)\times 10^{-7}$ 
has been also given by the LHCb~\cite{LHCb1}.
With $\bar B^0\to (c\bar c)(d\bar d)\to J/\psi p\bar p$, 
the $p\bar p$ production has the direct transition from $\bar B^0\to d\bar d \to p\bar p$, 
which associates with the threshold enhancement, such that theoretical prediction of 
$(11.4\pm 5.0)\times 10^{-7}$ in Refs.~\cite{BtoJpsippbar,BtoJpsippbar2} can be
consistent with the observation. On the contrary,
$\bar B^0_s\to J/\psi p\bar p$ via $\bar B^0_s\to s\bar s \to p\bar p$ 
leads to the OZI suppression, while $s\bar s$ should be first annihilated 
to produce $p\bar p$. 
With the OZI suppression of ${\cal B}(\phi\to \pi\pi)/{\cal B}(\phi\to K\bar K)\simeq 10^{-4}$ \cite{pdg},
one expects that ${\cal B}(\bar B^0_s\to J/\psi p\bar p)
\le 10^{-4} {\cal B}(\bar B^0_s\to J/\psi \Lambda\bar \Lambda)$, 
resulting in ${\cal B}(\bar B^0_s\to J/\psi p\bar p)\le10^{-9}$, while 
${\cal B}(\bar B^0_s\to J/\psi \Lambda\bar \Lambda)$ is considered to be at the same level as
${\cal B}(B^-\to J/\psi \Lambda\bar p)\simeq 1.18\times 10^{-5}$~\cite{pdg}.
Therefore, to understand the large branching ratio of around 
$3\times 10^{-6}$ for $\bar B^0_s\to J/\psi p\bar p$ in Eq.~(\ref{data1}),
a new theoretical study on this decay is clearly needed.

To explain ${\cal B}(\bar B^0_s\to J/\psi p\bar p)$, 
one possible solution is to have a resonant state
between the $s \bar s$ annihilation and  $p\bar p$ production in $\bar B^0_s\to J/\psi p\bar p$, so that
the process through its mass shell allows an on-shell enhancement for the decay branching ratio.
Indeed, it is common to observe resonant peaks in $B\to p\bar p M$. For example, one finds
the $c\bar c$ mesons, where the resonant $\eta_c\to p\bar p$ and $J/\psi\to p\bar p$ raise
the $m_{p\bar p}$ spectrum of $B^-\to K^- p \bar p$ \cite{Aaij:2013rha}, as well as
those identified as the charmed baryons and the glueball from $D^{(*)} p$ and $p\bar p$ spectra in
$\bar B^0\to D^{(*)0} p\bar p$~\cite{delAmoSanchez:2011gi,Hsiao:2013dta,Cheng:2012fq}, respectively.
According to Refs.~\cite{pdg,ppfJKK}, since $f_J(2220)\equiv f_J$ 
with the quantum numbers $J^{PC}=2^{++}$ or $4^{++}$ has the channel of $f_J\to p\bar p$, 
particularly, with its mass and decay width within the allowed region of the $m_{p\bar p}$ spectrum
in $\bar B^0_s\to J/\psi p\bar p$, it is reasonable that 
$f_J$ can be our candidate as the resonant state in $\bar B^0_s\to J/\psi p\bar p$.

The experimental status of $f_J$ is reported in Ref.~\cite{fJ_expt}, where
its evidences come from the Mark~III collaboration~\cite{MarkIII}
and the BES collaboration~\cite{BES}, also being supported by 
$\pi^-(K^-)p$ collisions~\cite{piKp}. However, the direct confirmations 
from $p\bar{p}$ collisions~\cite{pbarp} and  
$2\gamma$ processes~\cite{2gamma} are inconclusive.
Hence, it leaves the room for the $\bar B^0_s$ meson decays to provide 
the new scenario for the $f_J$ study. Moreover, 
according to the QCD models~\cite{fJ_thery}, 
such as the Lattice QCD (LQCD) calculation~\cite{LQCD},   
$f_J$ has the mass close to that of the tensor glueball ($G_{2^{++}}$) 
with $J^{PC}=2^{++}$.
Moreover,
the theoretical prediction of ~\cite{G2}
\begin{eqnarray}\label{G2_thery}
{\cal B}(J/\psi\to \gamma G_{2^{++}})=(1.1\pm 0.2\pm 0.1) \times 10^{-2}
\end{eqnarray}
agrees with the lower bound of 
${\cal B}(J/\psi\to \gamma f_J)>2.5\times 10^{-3}$~\cite{pdg}.
With $f_J(2220)$ being identified as $G_{2^{++}}$,
Eq. (\ref{G2_thery}) can be related to the radiative $J/\psi$ decays 
by the BABAR collaboration~\cite{fJ_BABAR}, 
given by~\cite{pdg}, 
\begin{eqnarray}\label{Jpsidecays}
&&{\cal B}(J/\psi\to \gamma f_J){\cal B}(f_J\to p\bar p,\,\pi\pi)=(1.5\pm 0.8,\,8\pm 5)\times 10^{-5}\,,\nonumber\\
&&{\cal B}(J/\psi\to \gamma f_J){\cal B}(f_J\to K\bar K)<3.6\times 10^{-5}\,,
\end{eqnarray}
such that we obtain
\begin{eqnarray}\label{fJdecays}
&&{\cal B}(f_J\to p\bar p,\,\pi\pi)=(1.4\pm 0.8,\,7.3\pm 3.9)\times 10^{-3}\,,\nonumber\\
&&{\cal B}(f_J\to K\bar K)<4.1\times 10^{-3}\,,
\end{eqnarray}
where the limit is based on the $1\sigma$ error of the measured value on $J/\psi\to \gamma G_{2^{++}}$.
We remark that the results in Eq.~(\ref{fJdecays}) 
are consistent with the ratios:
${\cal B}(f_J\to p\bar p,\,\pi\pi)/{\cal B}(f_J\to K\bar K)=(0.17\pm 0.09,\,1.0\pm 0.5)$ in the PDG~\cite{pdg}.

In this paper, we shall explain $\bar B^0_s\to J/\psi p\bar p$
with $f_J(2220)$ as the resonant state to $p\bar p$.
Due to this resonant state, we will also study the other hadronic decays of $\bar B^0_s$, such as 
 $\bar B^0_s\to J/\psi (\pi\pi\,, K\bar{K}\,, \Lambda\bar \Lambda)$ and
$\bar B^0_s\to D^{0*} (\pi\pi\,, K\bar{K}\,, p\bar p\,, \Lambda\bar \Lambda)$.

\section{Formalism}
\begin{figure}[t!]
\centering
\includegraphics[width=2.8in]{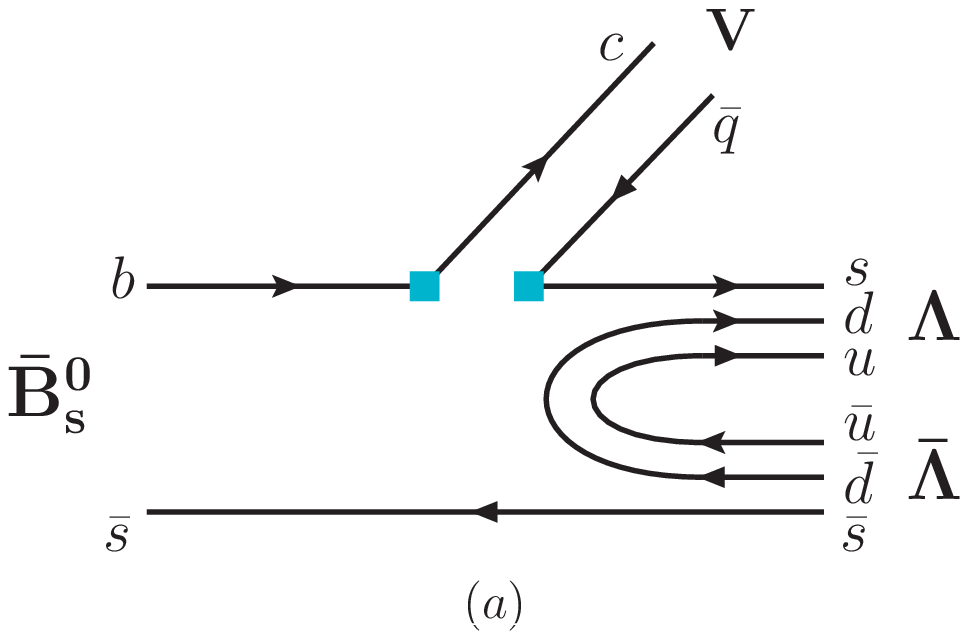}
\includegraphics[width=2.8in]{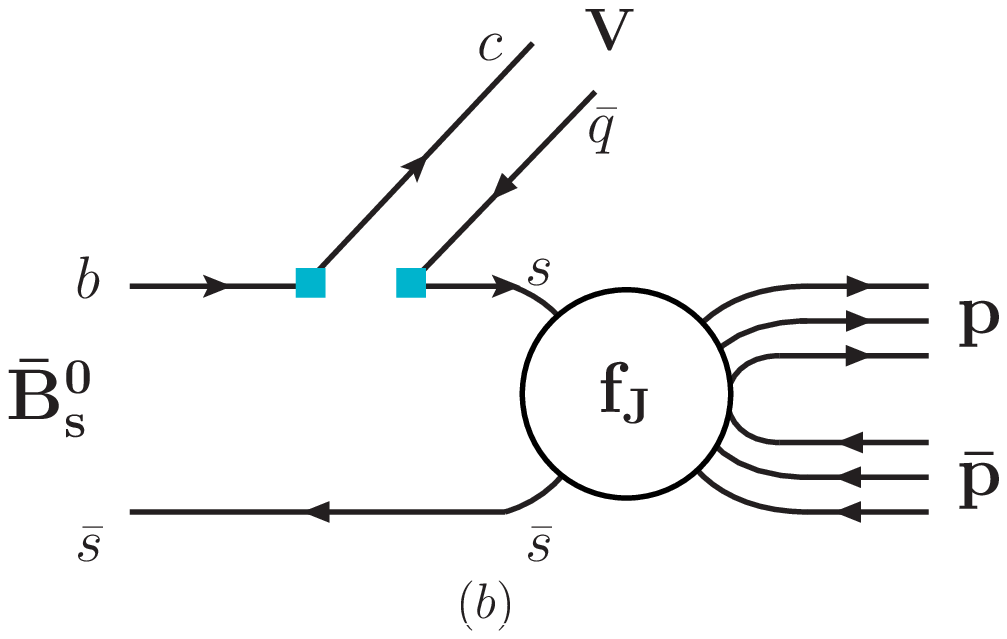}
\caption{The decays of $\bar B^0_s\to V {\bf B\bar B'}$ with 
${\bf B\bar B'}=$ (a) $\Lambda\bar \Lambda$ and (b) $p\bar p$,
produced by the pQCD effect and the resonance $f_J$, respectively,
where the block pairs represent the integrated-over $W$ boson 
in the effective Hamiltonian.  }\label{figVBB}
\end{figure}
%
In the effective Hamiltonian~\cite{Heff}, the amplitude of $\bar B^0_s\to V{\bf B\bar B'}$ 
with the baryon pair ${\bf B\bar B'}=p\bar p$ or $\Lambda\bar \Lambda$ can be factorized as
\begin{eqnarray}\label{amp1}
{\cal A}(\bar B^0_s\to V{\bf B\bar B'})=\frac{G_F}{\sqrt 2}V_{cb}V_{qs}^* a_2^{V}
\langle V|(\bar c q)_{V-A}|0\rangle\langle{\bf B\bar B'}|(\bar s b)_{V-A}|\bar B^0_s\rangle\,,
\end{eqnarray}
where $G_F$ is the Fermi constant, $V_{q_1q_2}$ are 
the Cabibbo-Kobayashi-Maskawa (CKM) matrix elements, 
and $a_2^V$ is the coefficient studied in Ref.~\cite{BtoJpsippbar},
while $(\bar q_1 q_2)_{V-A}$ denotes $\bar q_1\gamma_\mu(1-\gamma_5) q_2$
and $V$ stands for the vector meson $J/\psi$($D^{*0}$) with $q=c(u)$. 
In Eq. (\ref{amp1}), the matrix element of the vector meson production is defined by
\begin{eqnarray}\label{vectordecay}
\langle V|(\bar q c)_{V-A}|0\rangle&=&m_{V}f_{V}\varepsilon_\mu^*\,,
\end{eqnarray}
where $m_V$, $f_V$ and  $\varepsilon_\mu^*$ are the mass,  decay constant 
and polarization of the vector meson $V$, respectively.

For $\bar B^0_s\to V\Lambda\bar \Lambda$,
since the $s\bar s$ pair can have a direct transition to be
a part of the internal quarks in $\Lambda\bar \Lambda$ as seen in Fig.~\ref{figVBB}a, 
the most general matrix elements 
of the $\bar B^0_s\to \Lambda\bar \Lambda$ transition are given by~\cite{angulardistribution}
\begin{eqnarray}\label{transitionF}
&&\langle \Lambda\bar \Lambda|(\bar s b)_V|\bar B^0_s\rangle=
i\bar u[  g_1\gamma_{\mu}+g_2i\sigma_{\mu\nu}p^\nu +g_3p_{\mu} 
+g_4(p_{\bar \Lambda}+p_{\Lambda})_\mu +g_5(p_{\bar \Lambda}-p_{\Lambda})_\mu]\gamma_5v\,,\nonumber\\
&&\langle \Lambda\bar \Lambda|(\bar s b)_A|\bar B^0_s\rangle=
i\bar u[ f_1\gamma_{\mu}+f_2i\sigma_{\mu\nu}p^\nu +f_3p_{\mu} 
+f_4(p_{\bar \Lambda}+p_{\Lambda})_\mu +f_5(p_{\bar \Lambda}-p_{\Lambda})_\mu] v\,,
\end{eqnarray}
with $p=p_{\bar B^0_s}-p_{\Lambda}-p_{\bar\Lambda}$ 
and the form factors $g_i$ and $f_i$ (i=1,2, ..., 5).
In the approach of pQCD counting rules~\cite{countingrules1,countingrules2}, we are able to
count the number of the hard gluon propagators within the baryon pair, such that
the momentum dependences of $g_i$ and $f_i$ can be parameterized as~\cite{angulardistribution}
\begin{eqnarray}\label{transitionF2}
f_i=\frac{D_{f_i}}{t^n}\;, \qquad g_i=\frac{D_{g_i}}{t^n}\;,
\end{eqnarray}
with $t\equiv m_{\Lambda\bar \Lambda}^2=(p_{\Lambda}+p_{\bar\Lambda})^2$.
To the leading order, the counting gives $n=3$, 
in which 2 of them are for the gluons connecting to the valence quarks, 
while the rest one for the gluon speeding up $\bar s$ in $\bar B^0_s$ to be part of $\bar \Lambda$.
As $t$ approaches the threshold area, 
the increasing value of $1/t^3$ creates a peak in the $m_{\bf B\bar B'}$ spectrum of $B\to {\bf B\bar B'}$,
which interprets the threshold enhancement.
Under the SU(3) flavor symmetry, 
$D_{g_i}$ and $D_{f_i}$ are related 
by
$D_{g_1(f_1)}=D_{||}$,
and $D_{g_k}=-D_{f_k}=D^k_{||}$ ($k=2, 3,\cdots, 5$),
in which 
 the reduced constants $D_{||}$, ($D_{\overline{||}}$) and $D^k_{||}$
 can be fitted through the measured baryonic decays~\cite{BtoJpsippbar}. 

For $\bar B^0_s\to V p\bar p$, because
the matrix elements of  the $\bar B^0_s\to (s\bar s\to)p\bar p$ transition 
need the $s\bar s$ annihilation to produce $p\bar p$, 
which encounters the OZI suppression, it is not suitable for pQCD counting rules.
Consequently,
${\cal B}(\bar B^0_s\to J/\psi p\bar p)$ is estimated to be smaller than  $10^{-9}$
as mentioned early.
On the other hand, for the resonant transition of $\bar B_s^0\to f_J\to p\bar p$ as shown in Fig.~\ref{figVBB}b,
$m_{f_J}\simeq2.23$ GeV is in the $p\bar p$ invariant mass ($m_{p\bar p}$) spectrum, 
of which the range of 1.88 GeV$<m_{p\bar p}<$ 2.27 GeV is so confined,
such that the resonance has a complete peak, enhancing the decay branching ratio of 
$\bar B^0_s\to J/\psi p\bar p$.
The matrix element of the resonant $\bar B^0_s\to f_J\to p\bar p$ transition is given by
\begin{eqnarray}\label{BtofJtoppbar}
\langle p\bar p|(\bar s b)_{V-A}|\bar B^0_s\rangle=\langle p\bar p|f_J\rangle
\frac{i}{(t-m_{f_J}^2)+im_{f_J}\Gamma_{f_J}}\langle f_J|(\bar s b)_{V-A}|\bar B^0_s\rangle\,,
\end{eqnarray}
where   
$\Gamma_{f_J}$ ($m_{f_J}$) stands for the decay width (mass) of  $f_J$. 
%
In terms of 
Eqs. (\ref{amp1}), (\ref{vectordecay}), and (\ref{BtofJtoppbar}), 
we can write the amplitude of $\bar B^0_s\to V (f_J\to)p\bar p$ to be
\begin{eqnarray}\label{amp2}
{\cal A}_R(\bar B^0_s\to V(f_J\to) p\bar p)&=&\frac{G_F}{\sqrt 2}V_{cb}V_{qs}^*a_2^V
\frac{m_V f_V}{(t-m_{f_J}^2)+im_{f_J}\Gamma_{f_J}}\bar u(a+b\gamma_5)v\,,
\end{eqnarray}
with $\langle p\bar p|f_J\rangle$$\varepsilon^{\mu*} \langle f_J|(\bar s b)_{V-A}|\bar B^0_s\rangle$
$=\bar u(a+b\gamma_5)v$, where 
the the Lorentz indices
from four-momentum factors, $\varepsilon^{\mu*}$, 
and the summations of the spins for the intermediate $f_J$ state
are coupled to have a scalar quantity, leading $a$ and $b$ to be parameters.
Note that $a$ and $b$ are generally momentum dependent; however,
as the narrow range of  $m_{p\bar p}$ is 1.88-2.27 GeV due to the heavy $J/\psi$ mass,
 $a$ and $b$  can only be changed  slightly so that they are nearly constants. 
Moreover, the dominant contribution to the branching ratio
comes from the pole effect, which is even narrow, 
fixing the pole at $m_{f_J}=2.23$ GeV. 
In fact, $\langle p\bar p|f_J\rangle$ as the strong interaction conserves the parity, such that
it is in the form of either $\bar u v$ or $\bar u\gamma_5 v$ for the parity to be even or odd. 
Hence, while $f_J$ has been confirmed to have an even parity as the data indicated~\cite{pdg}, $b=0$.
In Eq.~(\ref{amp2}), since $a$ is unknown, 
it will be fitted with ${\cal B}(\bar B^0_s\to J/\psi p\bar p)$
and then used to predict ${\cal B}(\bar B^0_s\to D^{*0}p\bar p)$
as well as those of $\bar B^0_s\to J/\psi (\pi\pi,K\bar{K})$ and $\bar B^0_s\to D^{*0} (\pi\pi,K\bar{K})$. 
To integrate over the phase space of the three-body decays,
the general equation in the PDG~\cite{pdg} can be referred, which is given by 
\begin{eqnarray}\label{dG}
d\Gamma=\frac{1}{(2\pi)^3}\frac{|{\cal A}|^2}{32M^3_B}dm_{12}^2 dm_{23}^2\,,
\end{eqnarray}
where $m_{12}=p_{\bf B}+p_{\bf\bar B'}$, $m_{23}=p_{\bf B}+p_{V}$ and
$|{\cal A}|^2$ represents the amplitude squared. 
By integrating over the variables $m_{12}$ and $m_{23}$, we obtain
the total branching ratio. 
Here we will integrate over $m_{23}$ alone to have
the partial branching ratio as the function of $m_{\bf B\bar B'}$, 
such that the threshold and resonant effects 
drawn as the peaks in the $m_{\bf B\bar B'}$ spectra
would be in comparison with the future experiments.

\section{Numerical Results and Discussions}
\begin{figure}[t!]
\centering
\includegraphics[width=2.8in]{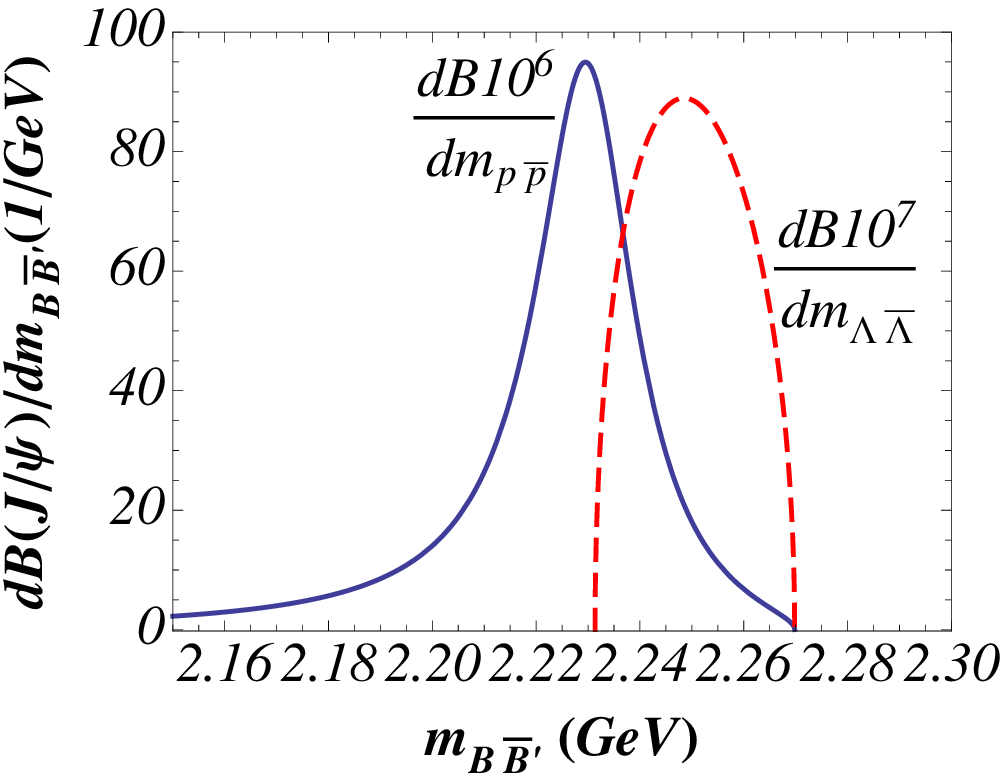}
\includegraphics[width=2.69in]{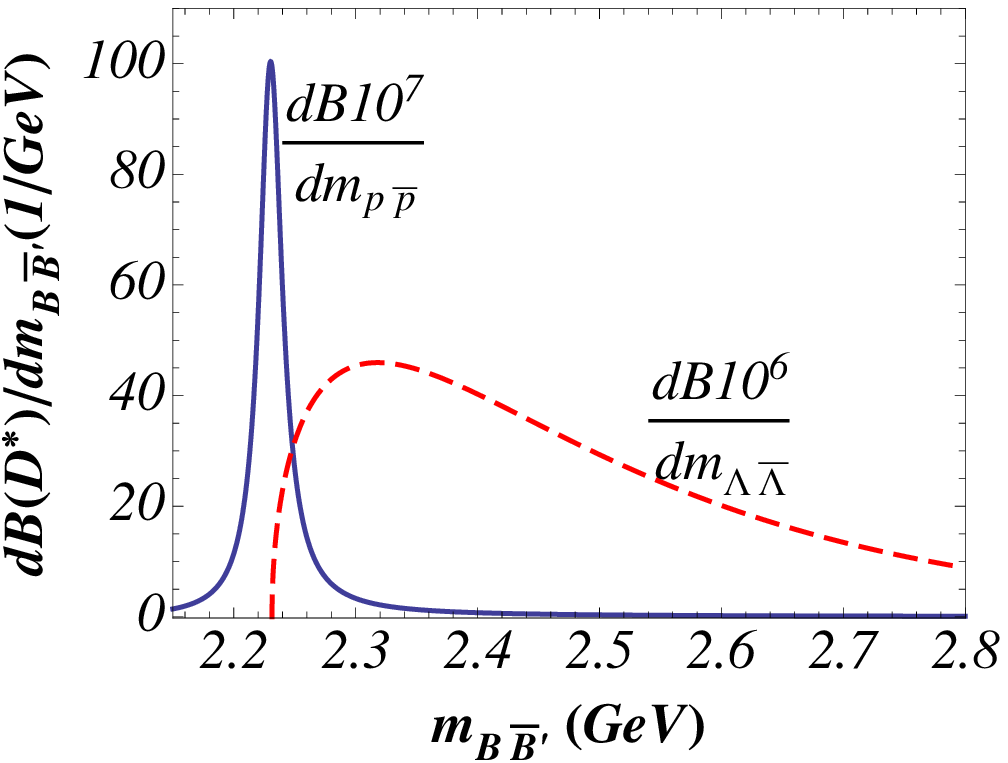}
\caption{The partial distributions vs. $m_{\bf B\bar B'}$ in
the $\bar B^0_s\to V {\bf B\bar B'}$ decays with $V=J/\psi(D^{*0})$, where
the left panel is for  $V=J/\psi$, 
while the right one for $V=D^{*0}$.
}\label{fig2VBB}
\end{figure}

In our numerical analysis, we adopt 
 ($m_{f_J}$, $\Gamma_{f_J}$)=(2231, 23) MeV and 
($V_{cb}$, $V_{cs}$, $V_{us}$)=($A\lambda^2$, $1-\lambda^2/2$, $\lambda$)
with $A=0.811$ and $\lambda=0.225$ in the PDG \cite{pdg}, and take
\begin{eqnarray}
&&(a_2^{D^*}, a_2^{J/\psi})=(0.33\pm 0.04,0.17\pm 0.03)\,,\nonumber\\
&&(D_{||},D_{\overline{||}})=(67.7\pm 16.3,-280.0\pm 35.9)\;{\rm GeV^5}\,,\nonumber\\ 
&&(D_{||}^2,D_{||}^3,D_{||}^4,D_{||}^5)=
(-187.3\pm 26.6,-840.1\pm 132.1,-10.1\pm 10.8,-157.0\pm 27.1)\;{\rm GeV^4},~~~
\end{eqnarray}
from the study of the charmful three-body baryonic $\bar B^0 (B^-)$ decays
in Ref.~\cite{BtoJpsippbar}.
For the decay constants, we use ($f_{D^*}$, $f_{J/\psi}$)=(0.23, 0.41) GeV~\cite{vectordecay}.
As a result, with 
$|a|$ fitted to be $1.04\pm 0.26$
we obtain ${\cal B}(\bar B^0_s\to J/\psi(f_J\to)p\bar p)=(3.00\pm 1.74)\times 10^{-6}$ to explain the data in Eq.~(\ref{data1}).
Consequently, we can calculate the branching ratios of 
$\bar B^0_s\to J/\psi \Lambda\bar \Lambda$ and
$\bar B^0_s\to D^{0*} p\bar p\,(\Lambda\bar \Lambda)$, 
of which the $m_{\bf B\bar B'}$ spectra are drawn in Fig. \ref{fig2VBB}, while 
the total branching ratios are listed in Table~\ref{branchingratios} with 
the errors coming from the uncertainties in various form factors.
\begin{table}[b!]
\caption{ The branching ratios of $\bar B^0_s\to V{\bf B\bar B'}$ decays, where
the uncertainties 
arise from $a_2^V$ and 
the $\bar B^0_s\to {\bf B \bar B'}$ transitions.
}\label{branchingratios}
\begin{tabular}{|c|c|}
\hline
decay mode               &branching ratio\\\hline
$\bar B^0_s\to  J/\psi (f_J\to)p\bar p$ 
& $(3.00\pm 1.74)\times 10^{-6}$\\
$\bar B^0_s\to D^{*0} (f_J\to)p\bar p$     
&$(4.70\pm 2.89)\times 10^{-7}$\\
$\bar B^0_s\to J/\psi \Lambda\bar \Lambda$ 
&$(2.68\pm 1.23)\times 10^{-7}$\\
$\bar B^0_s\to D^{*0}\Lambda\bar \Lambda$ 
&$(2.25\pm 0.80)\times 10^{-6}$\\
\hline
\end{tabular}
\end{table}
Since $\bar B^0_s\to J/\psi \Lambda\bar \Lambda$ and $B^-\to J/\psi \Lambda\bar p$
are essentially identical, except for the spectator quarks in $\bar B^0_s$ and $B^-$,
their branching ratios should be at the same level.
Nonetheless,  from Table~\ref{branchingratios}, we see that
${\cal B}(\bar B^0_s\to J/\psi \Lambda\bar \Lambda)\simeq 0.02
{\cal B}(B^-\to J/\psi \Lambda\bar p)$.
The reason for this is  that $m_{\Lambda\bar \Lambda}$ around the threshold area is smaller
than $m_{\Lambda\bar p}$ by 100 MeV,
which causes the more constrained threshold effect.
Similarly, for $D^{*0}$ cases, 
${\cal B}(\bar B^0_s\to D^{*0}\Lambda\bar \Lambda)\sim 2.25\times 10^{-6}$ is 
at least 20 times smaller than ${\cal B}(B^-\to D^{*0}\Lambda\bar p)$ \cite{DstarLpbar}.
%
%
It is interesting to note that the assumption of the constant
parameter of $a$ 
is demonstrated to be insensitive to the data fitting, while the pole effects 
via the resonant $f_J\to p\bar p$
in $\bar B^0_s\to J/\psi(D^{*0})p\bar p$ are narrow and sharp,
as shown in Fig.~\ref{fig2VBB}.
%
A further confirmation for the resonant $\bar B^0_s\to J/\psi (f_J\to)p\bar p$ 
can depend on the future search for $\bar B^0_s\to D^{*0}(f_J\to)p\bar p$,
whose decay branching ratio is predicted to be 
$(4.70\pm 2.89)\times 10^{-7}$ 
(see Table \ref{branchingratios}).
%
The difference between the threshold effect and the resonant $f_J$ peak can be seen from
Fig.~\ref{fig2VBB}, where the peaks from the threshold effects for 
$\bar B^0_s\to J/\psi(D^{*0})\Lambda\bar \Lambda$ are drawn to be smooth, whereas  the peaks
from the $f_J$ resonance are sharp with the highest point precisely at $m_{p\bar p}=2.23$ GeV
for $\bar B^0_s\to J/\psi(D^{*0})(f_J\to)p\bar p$.
This can be used for the future experiments to distinguish the threshold effect from the resonance.
Except for $\bar B^0\to J/\psi p\bar p$ with 1.88 GeV$<m_{p\bar p}<$ 2.18 GeV away from $m_J=2.23$~GeV,
the resonant contributions are also possible for the other $\bar B^0$ and $B^-$ decays of $B\to p\bar p M$, 
such as $B\to p\bar p K$ and $B\to p\bar p K^*$~\cite{Hou1}. Nonetheless,
the ratios of \cite{pdg,ppK_belle}
\begin{eqnarray}
&&{\cal B}(B^-\to K^-(f_J\to)p\bar p)/{\cal B}(B^-\to K^-p\bar p)<0.06-0.08\,,\nonumber\\
&&{\cal B}(B^-\to K^{*-}(f_J\to)p\bar p)/{\cal B}(B^-\to K^{*-}p\bar p)<0.18-0.27\,,\nonumber\\
&&{\cal B}(\bar B^0\to \bar K^0(f_J\to)p\bar p)/{\cal B}(\bar B^0\to \bar K^0 p\bar p)<0.15-0.19\,,\nonumber\\
&&{\cal B}(\bar B^0\to \bar K^{*0}(f_J\to)p\bar p)/{\cal B}(\bar B^0\to \bar K^{*0}p\bar p)<0.10-0.15\,,
\end{eqnarray}
are too small to have impacts on the experimental results,
due to the fact that the threshold effects in the decays shadow the resonant peaks.
%
Instead, 
the unexpected large value of ${\cal B}(\bar B^0_s\to J/\psi p\bar p)$ in Eq.~(\ref{data1})
would reveal the existence of $f_J(2220)$ due to the suppressed threshold effect in the decay.

In terms of ${\cal B}(\bar B^0_s\to V(f_J\to)AB)={\cal B}(\bar B^0_s\to V f_J){\cal B}(f_J\to AB)$
with ${\cal B}(\bar B^0_s\to V(f_J\to)AB)$ from Table~\ref{branchingratios}
and the $f_J\to AB$ decays in Eq. (\ref{fJdecays}), 
where $AB$ can be $p\bar p,\,K\bar K$, and $\pi\pi$,
we obtain
\begin{eqnarray}
&&{\cal B}(\bar B^0_s\to J/\psi f_J)=(2.1\pm 1.7)\times 10^{-3}\,,\nonumber\\
&&{\cal B}(\bar B^0_s\to J/\psi(f_J\to)K\bar K)<1.6\times 10^{-5}\,,\nonumber\\
&&{\cal B}(\bar B^0_s\to J/\psi(f_J\to)\pi\pi)=(15.6\pm 15.2)\times 10^{-6}\,,
\end{eqnarray}
for  $V= J/\psi$ and 
\begin{eqnarray}
&&{\cal B}(\bar B^0_s\to D^{*0} f_J)=(3.4\pm 2.8)\times 10^{-4}\,,\nonumber\\
&&{\cal B}(\bar B^0_s\to D^{*0}(f_J\to)K\bar K)<2.5\times 10^{-6}\,,\nonumber\\
&&{\cal B}(\bar B^0_s\to D^{*0}(f_J\to)\pi\pi)=(24.5\pm 24.4)\times 10^{-7}\,,
\end{eqnarray}
for $V= D^{*0}$. We then let the $\bar B^0_s$ decays be the new scenario to study the $f_J$ state.

\section{Conclusions}
We have studied  the roles of $f_J(2220)$,  
considered as the tensor glubeball state of $G_{2^{++}}$, 
in the hadronic $\bar B^0_s$ decays.
Explicitly, we have shown that the recent measured large  branching ratio  by the LHCb
for the  OZI suppressed decay of  ${\cal B}(\bar B^0_s\to J/\psi p\bar p)$  
can be understood due to the resonant contribution of $f_J(2220)$. We have also found that
${\cal B}(\bar B^0_s\to D^{*0}(f_J\to) p\bar p)=(4.70\pm 2.89)\times 10^{-7}$.
Similarly, we have predicted that
${\cal B}(\bar B^0_s\to J/\psi(f_J\to)\pi\pi)=(15.6\pm 15.2)\times 10^{-6}$ and 
${\cal B}(\bar B^0_s\to D^{*0}(f_J\to)\pi\pi)=(24.5\pm 24.4)\times 10^{-7}$,
while
${\cal B}(\bar B^0_s\to J/\psi(f_J\to)K\bar K)<1.6\times 10^{-5}$ and
${\cal B}(\bar B^0_s\to D^{*0}(f_J\to)K\bar K)<2.5\times 10^{-6}$.
In addition, we have obtained  that  ${\cal B}(\bar B^0_s\to (J/\psi\,,D^{*0})\Lambda\bar \Lambda)$ are
 $(2.68\pm1.23)\times 10^{-7}$ and $(2.25\pm0.80)\times 10^{-6}$, which
are accessible to the experiments at the LHCb.

\section*{ACKNOWLEDGMENTS}
This work was partially supported by National Center for Theoretical
Sciences,  National Science Council  
 (NSC-101-2112-M-007-006-MY3) and National Tsing Hua
University~(103N2724E1).


\begin{thebibliography}{99}
\bibitem{BtoJpsippbar2}
 The result in Ref.~\cite{BtoJpsippbar} has been updated  
according to the observed value of ${\cal B}(B^-\to J/\psi \Lambda\bar p)$.
\end{thebibliography}
\end{document}